\documentclass[12pt,epsf]{article}

\usepackage{amsmath,amssymb,graphicx}

\newcommand{\beq}{\begin{eqnarray}}
\newcommand{\eeq}{\end{eqnarray}}

\begin{document}

\begin{titlepage}
\vspace*{0.5cm}

\begin{center}
{\huge \bf A Quirky Little Higgs Model} \vspace{.3cm}

\end{center}
\vspace*{0.2cm}

\begin{center}
{\bf Haiying Cai, Hsin-Chia Cheng, and John Terning}

\end{center}
\vskip 8pt

\begin{center}{\it Department of Physics, University of California, Davis,
CA
95616.} \\
\vspace*{0.3cm}
\end{center}

\vglue 0.3truecm

\begin{abstract}

\vskip 3pt

 We consider an extra dimensional model  where the quadratically  divergent top loop contribution to the
Higgs mass is cancelled by an uncolored heavy ``top quirk" charged under a different $SU(3)$ gauge group.  The cancellation is enforced by bulk gauge symmetries. Thus we have an unusual type of little Higgs model which has some quirky signatures.  The top partner in this model could be identified at the Large Hadron Collider due to macroscopic strings that connect quirk and anti-quirks.  The model can undergo radiative electroweak symmetry breaking and is consistent with precision electroweak measurements.

\end{abstract}
\end{titlepage}

\newpage

\section{Introduction}

With new data from the Large Hadron Collider (LHC) imminent it would be ideal if all possible solutions for the little hierarchy problem of the Standard Model (SM)~\cite{Barbieri:2000gf} had been classified and their LHC phenomenology at least qualitatively explored.   Understandably particle theorists have not been equal to this monumental task,
and new proposals continue to trickle in.  There has also recently been some interest in exploring what the LHC could possibly find that is not directly motivated by solving the little hierarchy problem.  Examples of these ``unmotivated models" include hidden valleys~\cite{Strassler}, quirks~\cite{quirk}, and unparticles~\cite{Georgi}.  If such ``unmotivated" physics was uncovered at the LHC one would have to wonder: ``how does it fit into the solution of the little hierarchy problem?"  One possibility is that such new physics is part of some sector unrelated to electroweak symmetry breaking that just happens to have a similar mass threshold which  might have its own hierarchy problem.  Here we will explore the more intriguing possibility that the ``unmotivated" physics turns out to  actually be part of the solution rather than part of a new problem.

In the quirk scenario~\cite{quirk} there are some new fermions that couple to  a new non-Abelian
gauge group referred to as infracolor. The fermions charged under this new gauge group are called quirks in analogy to the traditional quarks.  The quirks may or may not be coupled to some or all of the SM gauge groups.
How could quirks solve the little hierarchy problem? Recall that in the folded
SUSY model~\cite{FD}, the quadratic divergence from the top quark is cancelled by its folded-partner, which is a spin-0 scalar not charged
under the SM $SU(3)_C$ color gauge group. From the model building perspective, one sees that if we want the quadratic divergence from the top quark
to be cancelled by a spin-1/2 partners as in the Little Higgs mechanism, then the new particles are not necessarily colored as in the original little
Higgs model~\cite{ArkaniHamed:2001nc,LH}. However, without a symmetry reason that requires the cancellation of the divergence, this would not amount to a solution, but just a different form of fine-tuning.  Folded-SUSY is one possible symmetry for ensuring a cancellation between seemingly unrelated particles.  Here we will see that embedding the color and infracolor gauge groups in a larger gauge symmetry can also ensure a cancellation between top quark loops and ``top quirk" loops in the Higgs mass calculation.

 In section 2, we present the
details of our five dimensional model, including the embedding of color and infracolor in a larger bulk gauge group. Section 3 deals with the extended electroweak gauge group $SU(3)_W \times U(1)_X$ that is needed for the little Higgs mechanism. The extended gauge groups are broken to the SM by
boundary conditions. Section 4 discusses the mass spectrum and how it is  modified by boundary terms. In section 5, we calculate the one-loop corrections to
the Higgs mass parameter in momentum-position space. Section 6 is devoted to discussing the precision electroweak constraints on our model
parameters. The oblique parameters $S, T$ as well as $Z \to b_L \bar b_L$ are calculated and bounds on the size of the extra dimension are extracted.
In section 7, we discuss the experimental signatures of our top quirks in different cases. We also calculate the pair production rate for top quirks at LHC.
Our model is similar to the recently proposed dark top model~\cite{dark}. The dark top model also has a $SU(6)$ symmetry which relates the top quark with
its fermion partner. However in that model the  top quarks and top partners are put in special incomplete $SU(6)$ representations. Another difference is that
only a $SU(3)_C$ subgroup in the $SU(6)$ global symmetry group is gauged. The top partners in Ref.~\cite{dark} are gauge neutral so that they  can be identified as
dark matter.

\section{The model}

In this paper  we consider a new variation of the little Higgs mechanism for solving the little hierarchy problem. Because the heavy top partner in this model is
not colored but transforms under a new non-Abelian gauge group, we will call it the top quirk.  The model is five dimensional; the extra
dimension has a radius $R$ and is orbifolded by $S^1/Z_2$. The two boundaries are located at $y=0$ and $y= \pi R$. The bulk preserves an $SU(6) \times
SU(3)_W \times U(1)_X$ gauge symmetry. The $SU(6)$ bulk gauge symmetry relates the top quark to its partner top quirk. The $SU(3)_W$ bulk gauge
symmetry plays the same role as the $SU(3)_W$ gauge symmetry in the original littlest Higgs model~\cite{LH}. The $SU(3)_W$ gauge symmetry is broken twice, by the boundary condition at the $y=0$ boundary and also by the vacuum expectation value (VEV) of a scalar field $\Phi$ which transforms as a fundamental representation of $SU(3)_W$, living on the $y=\pi R$ boundary.  Since the collective symmetry breaking is nonlocal in this
model, the one-loop corrections to the Higgs mass parameter will be insensitive to the UV cutoff. The Higgs doublet is the pseudo-Nambu-Goldstone boson (PNGB) of the broken $SU(3)_W$ symmetry and effectively lives on the $y=\pi R$ boundary. The $y=0$ boundary preserves only $SU(3)_C \times SU(3)_I \times SU(2)_W \times U(1)_Y$ gauge symmetry, where $SU(3)_C \times SU(3)_I \subset SU(6)$ and $SU(2)_W \times U(1)_Y \subset SU(3)_W \times U(1)_X$. On the other hand, the $y=\pi R$ boundary preserves the whole $SU(6)\times SU(3)_W \times U(1)_X$ symmetry before the scalar VEV is turned on.

As for the SM fermion charge assignments, let us consider the third generation first. We put the left-handed top and bottom quarks in the bulk. They are embedded in ${\bf Q}_{t_L}^{(6,3)}$, which transforms as a
bi-fundamental of $SU(6)\times SU(3)_W$:
\[SU(6) \supset SU(3)_C \times SU(3)_I\]
\begin{eqnarray}
\begin{array}{clr} \mathbf{Q}_{t_L}^{\mathbf{(6,3)}}=(Q_{t_L},\, Q_{T_L}): & \left\{
\begin{array}{cc}
   {t_L ( ++ )} & {T_L ( -+ )}  \\
   {b_L ( ++ )} & {B_L ( -+ )}  \\
   {\chi ( -+ )} & {X( ++ )}  \\
\end{array}  \right \} & \left\{
\begin{array}{cc}
   {t^c_L ( -- )} & {T^c_L ( +- )}  \\
   {b^c_L ( -- )} & {B^c_L ( +- )}  \\
   {\chi^c ( +- )} & {X^c( -- )}  \\
\end{array}\right \}
\end{array}
\end{eqnarray}
The capital letters denote fields which transform under $SU(3)_I$. The fields with a superscript ``c'' are the chirality partner fermions in 5 dimensions and they carry the conjugate gauge quantum numbers.
The boundary conditions at the $y=0$ and $y=\pi R$ are indicated in  parentheses. We see that only the top-bottom quark doublet and a ``top quirk'' $X$ which transforms under $SU(3)_I$ have zero modes.
In order to unify the Yukawa couplings of the top quark and the top quirk, the right-handed top quark is chosen to live on the $y=\pi R$ boundary. It is embedded in $\mathbf{t}_R^{(6,1)}= \{t_R, T_R \} $ so it also has a quirk partner $T_R$ which marries with $X$ by the $\Phi$ VEV.

The right-handed bottom quark, on the other hand, lives in the bulk, because we do not want any light quirk in the spectrum. The right-handed bottom  quark is embedded in $\mathbf{Q}_{b_R}^{(6,\bar{3})}$:
\begin{eqnarray}
\begin{array}{clr} \mathbf{Q}_{b_R}^{\mathbf{(6,\bar 3)}}=(Q_{b_R},\,Q_{B_R}): & \left\{
\begin{array}{cc}
   {\delta_R ( -+ )} & {\Delta_R ( ++ )}  \\
   {v_R ( -+ )} & {\Upsilon_R ( ++ )}  \\
   {b_R ( ++ )} & {B_R ( -+ )}  \\
\end{array} \right\}& \left\{
\begin{array}{cc}
   {\delta_R^c ( +- )} & {\Delta_R^c ( -- )}  \\
   {v_R^c ( +- )} & {\Upsilon_R^c ( -- )}  \\
   {b_R^c ( -- )} & {B_R^c ( +- )}  \\
\end{array} \right\}
\end{array}.
\end{eqnarray}
The zero modes of $\Delta_R$ and $\Upsilon_R$ can obtain a large mass by marrying fields living on the $y=0$ boundary with conjugate quantum
numbers and hence are removed from the low energy spectrum. Equivalently, we could choose their boundary conditions to be $(-+)$ so that there
are no exotic zero modes from the beginning.

The Yukawa interaction and mass term for the third generation are:
\begin{eqnarray}
{\mathcal L}_{5D Yuk.}&=&
\lambda _t \Phi^{\dag (1,\bar 3)} \mathbf{Q}_{t_L}^{(6,3)} \bar{ \mathbf{t}}_R^{( \bar 6,1)}
\delta (y - \pi R) + \lambda_b \Phi^{(1, 3)} \mathbf{Q}_{t_L}^{(6,3)} \bar{ \mathbf{Q}}_{b_R}^{(
\bar 6, 3)} \delta (y - \pi R) \nonumber\\ && +
M_B (\Delta_L^{\prime}\bar \Delta_R+ \Upsilon_L^{\prime} \bar \Upsilon_R) \delta (y),
\end{eqnarray}
where $(\Delta'_L,\, \Upsilon'_L)$ is an $SU(2)_W$ doublet living on $y=0$ boundary which lifts the zero modes of $\Delta_R,\, \Upsilon_R$.

The $SU(3)_W$ gauge symmetry is broken by the boundary condition at $y=0$ and also by the VEV of $\Phi$ at the $y=\pi R$ boundary.  The Higgs field is the uneaten PNGB which
lives in $\Phi$ as long as $\langle \Phi \rangle =f \ll R^{-1}$. In the nonlinear sigma model notation, $\Phi$ can be expanded as
\begin{eqnarray}
\begin{array}{l}
 \Phi  = \left( {\begin{array}{c}
   {\bf 0}  \\
   {f}  \\
\end{array}} \right) + i \left( {\begin{array}{c}
   H  \\
   0  \\
\end{array}} \right) - \frac{1}{{2 f }}\left( {\begin{array}{c}
   {\bf 0}  \\
   {H^\dag  H}  \\
\end{array}} \right) + \cdots .
 \end{array}
\end{eqnarray}
When we expand the bulk top Yukawa interaction in component fields and only consider the zero mode contributions, we get the following
expression:
\begin{eqnarray}
{\mathcal L}_{Yuk.}&=&
 i\lambda _t H t_L t_R  + \lambda _t f X^c T_R^c
 - \frac{\lambda _t }{{2 f }} X^c T_R^c H^\dagger H .\label{bulk}
\end{eqnarray}
We can see that at the one-loop level the quadratically divergent contribution to the Higgs mass-squared from the top quarks $t_L, t_R$ is cancelled by that
from the heavy top quirks $X^c, T_R^c$, which are uncolored and charged under the $SU(3)_I$ gauge group. The necessary relation between the couplings is enforced by the bulk $SU(6)$ gauge symmetry.

In addition to the Higgs, there is a SM singlet PNGB $\eta$ which also receives its mass from nonlocal gauge loops. It only couples to SM fermions through their mixings with heavy fermions so it interacts with SM fields very weakly. It does not play any important role in phenomenology~\cite{Berezhiani:2005pb}.

We assign the SM fermions of the two light generations to the bulk. The left-handed quark doublets and
the right-handed down-type quarks are embedded in the same way as the third generation. The right-handed up-type
quarks also live in the bulk to avoid having very light quirks associated with the first two generations.
\begin{eqnarray}
\begin{array}{clr} \mathbf{Q}_{u_L}^{i \mathbf{(6,3)}}: & \left\{
\begin{array}{cc}
   {u_L^i ( ++ )} & {U_L^i ( -+ )}  \\
   {d_L^i ( ++ )} & {D_L^i ( -+ )}  \\
   {\chi_u^i ( -+ )} & {X_U^i( ++ )}  \\
\end{array}\right\} & \left\{
\begin{array}{cc}
   {u^{i c}_L ( -- )} & {T^{i c}_L ( +- )}  \\
   {d^{i c}_L ( -- )} & {B^{i c}_L ( +- )}  \\
   {\chi^{i c}_u ( +- )} & {X^{i c}_U( -- )}  \\
\end{array} \right \}
\end{array}
\end{eqnarray}
\begin{eqnarray}
\begin{array}{clr}\mathbf{Q}_{d_R}^{\mathbf{(6,\bar 3)}}: & \left\{
\begin{array}{cc}
   {\delta_{d R}^i ( -+ )} & {\Delta_{d R}^i ( ++ )}  \\
   {v_{d R}^i ( -+ )} & {\Upsilon_{d R}^i ( ++ )}  \\
   {d^i_R ( ++ )} & {D^i_R ( -+ )}  \\
\end{array} \right \} & \left\{
\begin{array}{cc}
   {\delta_{d R}^{i c} ( +- )} & {\Delta_{d R}^{i c} ( -- )}  \\
   {v_{d R}^{i c} ( +- )} & {\Upsilon_{d R}^{i c} ( -- )}  \\
   {d_{R}^{i c} ( -- )} & {D_{R}^{i c} ( +- )}  \\
\end{array} \right \}
\end{array}
\end{eqnarray}
\begin{eqnarray}
\begin{array}{clr} \mathbf{u}_R^{i \mathbf{(6,1)}} : &
\begin{array}{cc}
{u_R^i ( ++ )} & {U_R^i ( -+ )}  \\
\end{array} &
\begin{array}{cc}
{u^{c , i}_R ( -- )} & {U^{c , i}_R ( +- )}  \\
\end{array}
\end{array}
\end{eqnarray}
 The Yukawa interactions for the light generations are:
\begin{eqnarray}
&& \lambda _u \cdot \Phi^{\dag (1,\bar 3)} \mathbf{Q}_{u_L}^{i (6,3)} \bar{\mathbf{ u}}_R^{i (\bar 6,1)} \cdot \delta (y - \pi R) + \lambda_d \cdot \Phi^{(1, 3)}
\mathbf{Q}_{u_L}^{i (6,3)} \bar{\mathbf Q}_{d_R}^{i (\bar
6, 3)} \cdot \delta (y - \pi R) \nonumber \\
&&+ M_X^i \cdot X_U^{\prime i c}\bar{X}_U^{i} \cdot \delta (y) + M_D^i \cdot (\Delta_{d L}^{i \prime} \bar \Delta_{d R}^i+ \Upsilon_{d L}^{i \prime} \bar \Upsilon_{d
R}^i) \cdot \delta (y),
\end{eqnarray}
where $X_U^{\prime i c}, \, \Delta_{d L}^{i \prime},\ \Upsilon_{d L}^{i \prime}$ are fields living on $y=0$ boundary which lift the extra zero modes of $X_U^i,\, \Delta_{d R}^i,\, \Upsilon_{d R}^i$.

\section{Gauge Fields and Hypercharge}

As we mentioned earlier, the bulk $SU(6)$  gauge symmetry is broken down to $SU(3)_C \times SU(3)_I$ by the boundary conditions at $y=0$ and as a result, quarks
are charged under the $SU(3)_C$ gauge group while quirks are charged under the $SU(3)_I$. The boundary conditions at $y=0$ also break the $SU(3)_W \times U(1)_X$ into
$SU(2)_L \times U(1)_Y$. In the orbifold language, we can assign odd parity to the 4-dimensional off-diagonal gauge fields when reflecting at $y=0$. For example, the orbifold parities at the $y=0$ and $y=\pi R$ fixed points of the $SU(3)_W$ gauge fields are
\[
\left( {\begin{array}{cc|c}
   {( +  + )} & {( +  + )} & {( -  + )}  \\
   {( +  + )} & {( +  + )} & {( -  + )}  \\ \hline
   {( -  + )} & {( -  + )} & {( +  + )}  \\
\end{array}} \right).
\]
As the orbifold symmetry breaking does not reduce the rank, the extra $U(1)$s can be removed by introducing large VEVs of correspondingly charged fields at the $y=0$ boundary fields. These fields can be decoupled~\cite{GA} in the limit  of infinite VEVs and the Dirichlet boundary condition is recovered for the $U(1)$ gauge fields. For example, the $U(1)$ corresponding to the $T_8$ generator of the $SU(3)_W$ gauge group is not broken by the orbifold. However, it does not correspond to the correct hypercharge. To obtain the correct hypercharge gauge group, the extra $U(1)_X$ is needed. We can introduce a scalar field charged under both $T_8$ and $U(1)_X$ on the  $y=0$ boundary, with  a large VEV which breaks them down to the diagonal  subgroup. The unbroken linear combination of $T^8$ and
$U(1)_X$ which gives  the hypercharge $Y$ is:
\begin{eqnarray}\begin{array}{lcr}
Y= \frac{1}{\sqrt 3 }T^8 + X,  & \qquad \mbox{where} \qquad &   T^8  = \frac{1}{{\sqrt 3 }}(-\frac{1}{2},-\frac{1}{2}, 1).
\end{array}\end{eqnarray}
The $U(1)_X$ charge assignments for the Higgs and the third generation are shown in Table~\ref{hyper}, and the light generations are similar.

\begin{table}[!htb]\begin{center}\begin{tabular}{|c|c|c|c|c|c|} \hline & $\Phi^{(1,3)}$ & $\mathbf{Q}_{t_L}^{(6,3)}$ &
$\mathbf{t}_R^{(6, 1)}$ & $\mathbf{Q}_{b_R}^{(6,\bar 3)}$ & $(\Delta_L^{\prime}, \Upsilon_L^{\prime})$ \\\hline $U(1)_X$ &
$-\frac{1}{3}$ & $\frac{1}{3}$ & $\frac{2}{3}$ & $0$ & $0$
\\\hline
\end{tabular}\end{center} \caption{the $U(1)_X$ assignment for the scalars and top multiplets}\label{hyper}\end{table}

\section{The Mass Spectrum}

Before the VEV of $\Phi$ on the $y= \pi R$ boundary is turned on, the mass spectrum of the bulk fields as shown in Table~\ref{mass}
is determined by the boundary conditions.
\begin{table}[!htb]\begin{center}\begin{tabular}{|c|c|c|c|c|} \hline $(n=0, 1 ,2 \cdots)$ & $(++)$ & $(-+)$ &
$(+-)$ & $(--)$ \\\hline $m_n$ & $\frac{n}{R}$ & $\frac{n+1/2}{R}$ & $\frac{n+1/2}{R}$ & $\frac{n+1}{R}$
\\\hline
\end{tabular}\end{center} \caption{ mass spectrum under the boundary condition assignments} \label{mass}
\end{table}
After the third component of $\Phi$ gets a VEV $f$,  the back-reaction effects from
the boundary fields need to be taken into account and the mass spectrum of particles with positive boundary conditions will be modified. We can take $X_L(++)$ and $\chi(-+)$ as
examples to illustrate the modifications. Considering $X_L(++)$ first, the boundary Yukawa interaction $\lambda_t \Phi^{\dag} X_L(++) \bar T_R $ will
mix the original KK modes \cite{MM}. We can use the equations of motion and modified boundary conditions to obtain the new mass spectrum~\cite{FE}.

Expanding $X_L(++)$, its five dimensional partner $X_L ^c(--)$ and $T_R$ in KK modes:
\begin{eqnarray}
 X_L  &=& \sum\limits_n {g_n (y)\chi _n (x)},  \\
 \overline{ X_L ^c } &=& \sum\limits_n {f_n (y)\bar \psi _n (x)},  \\
 T_R  &=& \sum\limits_n {h_n \bar \psi _n (x)}.
 \end{eqnarray}
 The profile functions satisfy the following equations:
\begin{eqnarray}
&& \partial _5 f_n (y) + m_n g_n (y) - \lambda fh_n \delta (y - L)L^{1/2}  = 0, \\
&& \partial _5 g_n (y) - m_n f_n (y) = 0.
 \end{eqnarray}
We need to fix two boundary conditions of $f_n$ or $g_n$ at $y=0$ and $y= \pi R$, and the boundary condition at $y=\pi R$ is modified by the boundary Yukawa interaction. The two
boundary conditions are:
\begin{eqnarray}
&& \left. {f_n (y)} \right|_{y = 0}  = 0, \\
&& \left. {f_n (y)} \right|_{y = L}  =  - \lambda fL^{1/2} h_n.
 \end{eqnarray}
With the canonical normalization conditions,
\begin{eqnarray}
\begin{array}{lcr}
 \int_0^L dy f_n^2(y) + h_n^2 = 1, & \quad  &
 \int_0^L dy g_n^2(y)  = 1,
\end{array}
 \end{eqnarray}
we can solve the equations and obtain the mass spectrum:
\begin{eqnarray}
m_n \cdot \tan m_n L = \lambda ^2 f^2 L.
\end{eqnarray}

For the $\chi( -  + )$ case, the bulk functions are the same but the boundary conditions are different. Let the profile of $\chi(
-  + )$ be $f_n(y)$ and the profile of its five dimensional partner $\chi^c( + - )$  be $g_n$, we find
\begin{eqnarray}
&& \left. {g_n (y)} \right|_{y = 0}  = 0, \\
&& \left. {f_n (y)} \right|_{y = L}  =  - \lambda fL^{1/2} h_n,
 \end{eqnarray}
\begin{eqnarray}
\begin{array}{lcr}
 \int_0^L dy g_n^2(y) + h_n^2 = 1, & \quad &
 \int_0^L dy f_n^2(y)  = 1.
\end{array}
 \end{eqnarray}
 Similar to the  procedure for the $(++)$ case, the mass spectrum of the $(-+)$ is determined to be:
\begin{eqnarray}
m_n \cdot \cot m_n L = - \lambda ^2 f^2 L.
\end{eqnarray}

\section{The Higgs Potential}

 One way to calculate the one-loop radiative corrections to the scalar field is to sum over all the KK modes. However, since the
symmetry that protects the Higgs mass parameter is broken non-locally, it is more convenient to calculate the Higgs potential in the mixed
momentum-position space~\cite{MP}. The propagators for the bulk fields can be calculated by solving the equation: $(p^2-\partial_5^2) G (p, y,
y^{\prime}) = \delta (y-y^{\prime})$. Putting appropriate boundary conditions, we get:
\begin{eqnarray}
 \tilde G_p^{( +  + )} (y,y') &=& \pi R\frac{{(e^{py_ <  }  + e^{ - py_ <  } )(e^{py_ >  }  + e^{2p\pi R} e^{ - py_ >  } )}}{{2p(e^{2p\pi R}  - 1)}},
 \label{pp}  \\  \nonumber  \\
 \tilde G_p^{( -  + )} (y,y') &=& \pi R\frac{{(e^{py_ <  }  - e^{ - py_ <  } )(e^{py_ >  }  + e^{2p\pi R} e^{ - py_ >  } )}}{{2p(e^{2p\pi R}  +
 1)}}. \label{mp}
\end{eqnarray}
The five-dimensional couplings and four dimensional couplings are related by
\begin{equation}
 g_5 = \sqrt {\pi R} \cdot g_2, \qquad \qquad    g_{x5} = \sqrt {\pi R} \cdot g_x, \qquad \qquad  \lambda_t = \sqrt {\pi R} \cdot h_t .
\end{equation}

 The mass parameters can be calculated using the Coleman-Weinberg potential~\cite{RA}.  The contribution from the $SU(3)$ gauge fields is finite and positive:
\begin{eqnarray}
m^2 _{H}|_{\mbox{gauge}}  &=& \frac{{9g_2^2 }}{4}\int_0^\infty  {\frac{{p^3 dp}}{{8\pi ^2 }}}  \frac{2}{3}  \left[ {G_p^{( +  + )} (L,L) -
G_p^{(
- + )} (L,L)} \right] \nonumber \\
& = &  \frac{{3 g_2^2 }}{2} \int_0^\infty  \frac{p^2 dp}{8 \pi^2} \cdot \frac{{\pi R}}{p} \cdot \left( {\frac{{e^{2p\pi R}  + 1}}{{e^{2p\pi R} -
1}}
- \frac{{e^{2p\pi R}  - 1}}{{e^{2p\pi R}  + 1}}} \right) \nonumber \\
& = &  \frac{21 g_2^2 \zeta (3) }{128 \pi ^4 R^2}.
\end{eqnarray}
Here $G_p^{( +  + )} (L,L)$ and $G_p^{( -  + )} (L,L)$
are the 5D propagators evaluated on the $y= L$ boundary where the Higgs lives.
The contribution from the top quark triplet $Q_{t_L}$ is negative and the contribution from top quirk triplet $Q_{T_L}$ is positive:
\begin{eqnarray}
m_H^2 |_{Q_{t_L}}  &=& - 6 h_t^2  \int_0^\infty  {\frac{{p^3 dp}}{{8\pi ^2 }}\sum\limits_{n = 1}^\infty {G_p^{( -  + )(n - 1)}
(L,L)( - h_t^2 f^2 )^{n - 1} } \left[ {G_p^{( +  + )} (L,L) - G_p^{( -  + )} (L,L)} \right]}\nonumber \\
&=& - 6 h_t^2 \int_0^\infty  {\frac{{p^3 dp}}{{8\pi ^2 }} \frac{{G_p^{( +  + )} (L,L) - G_p^{( -  + )} (L,L)}}{{1 + G_p^{( -  + )} (L,L)h_t^2
f^2 }}}
\end{eqnarray}
\begin{eqnarray}
m^2 _{H}|_{Q_{T_L}} &=&  6 h_t^2 \int_{0}^\infty  {\frac{{p^3 dp}}{{8\pi ^2 }}} \sum\limits_{n = 1}^\infty{G_p^{( +  + )(n - 1)} (L,L)(-
h_t^2 f^2 )^{n - 1} } \left[
{G_p^{( +  + )} (L,L) - G_p^{( -  + )} (L,L)} \right] \nonumber \\
&=& 6 h _t^2 \int_0^\infty  {\frac{{p^3 dp}}{{8\pi ^2 }}\frac{{G_p^{( +  + )} (L,L) - G_p^{( -  + )} (L,L)}}{{1 + G_p^{( +  + )} (L,L) h_t^2 f^2
}}}
\end{eqnarray}

To calculate the contribution to the Higgs mass-squared, we need to sum over all the mass insertions in the Coleman-Weinberg potential and keep
only the $H^\dagger H$ term. The summing process actually gives a renormalization for all the pole masses in the propagator functions,  as we
discussed in the previous section, and the zero modes get obvious shifts. The denominators in the above two expressions effectively give an
infrared cut off for the integrals.  Combining the above two contributions, we have
\begin{eqnarray}
m_H^2 |_{Q_{t_L}} + m_H^2 |_{Q_{T_L}} &\simeq &  \frac{3}{4 \pi^2} \cdot h_t^4 \cdot f^2 \cdot \{\log (4 \pi  h_t f \cdot R)- ({2 \pi h_t f
\cdot R} + 1)\}
\end{eqnarray}
In order for to get radiative electroweak symmetry breaking, the magnitude of this contribution needs to be larger than the gauge contributions.
For $h_t \sim 1$, this roughly requires $f \cdot R > 0.084 $.

Radiative correction from the top quarks and quirks will contribute to the quartic term for the higgs fields. Expanding the higgs potential to
$H^{\dagger}HH^{\dagger}H$ order, we get:
\begin{eqnarray}
 \left. {\lambda _H } \right|_{Q_{tL} }  &=& 2 \cdot \frac{{h_t^2 }}{{f^2 }} \int_0^\infty  {\frac{p^3 dp}{8\pi ^2 } \frac{{G_p^{( +  + )} (L,L) - G_p^{( -  + )} (L,L)}}{{1 + G_p^{( -  + )} (L,L)h_t^2 f^2 }}} \nonumber \\
  &+& 3 \cdot h_t^4 \int_0^\infty  {\frac{p^3 dp}{8\pi ^2 } \left( {\frac{{G_p^{( +  + )} (L,L) - G_p^{( -  + )} (L,L)}}{{1 + G_p^{( -  + )} (L,L)h_t^2 f^2 }}}
  \right)^2}
\end{eqnarray}
\begin{eqnarray}
 \left. {\lambda _H } \right|_{Q_{TL} }  &=& 2 \cdot \frac{{h_t^2 }}{{f^2 }} \int_0^\infty  {\frac{p^3 dp}{8\pi ^2 } \frac{{G_p^{( -  + )} (L,L) - G_p^{( +  + )} (L,L)}}{{1 + G_p^{( +  + )} (L,L)h_t^2 f^2 }}} \nonumber \\
  &+& 3 \cdot h_t^4 \int_0^\infty  {\frac{p^3 dp}{8\pi ^2 } \left( {\frac{{G_p^{( -  + )} (L,L) - G_p^{( +  + )} (L,L)}}{{1 + G_p^{( +  + )} (L,L)h_t^2 f^2 }}}
  \right)^2}
\end{eqnarray}
\beq\left. {\lambda _H } \right|_{Q_{tL} } + \left. {\lambda _H } \right|_{Q_{TL} } \simeq   - \frac{1}{\pi^2} \cdot h_t^4 \cdot \{\log (4 \pi
h_t f \cdot R )- ({2 \pi h_t f \cdot R} + 1)\}\eeq
For $h_t \sim 1$, we find $\lambda_H > 0.12 $, where the minimal value occurs at $f \cdot R \simeq 0.16$, this translates to a physical mass for
the higgs bosons $m_h > 121 ~ \mbox{GeV}$, which is larger than the LEP $2$ bound.

\section{Electroweak Precision Constraint}

 The size of the extra dimension is constrained by precision  electroweak measurements. Oblique corrections to the Standard Model contained in the vacuum polarizations  of gauge bosons, which are parameterized by $S$ and $T$~\cite{EW,TC}. Since the third generation is treated differently from the others, constraints from
$Z \rightarrow b_L b_L $ also need to be considered.

The oblique parameters $S, T$ are related to electroweak symmetry breaking. $S$ roughly measure the size of the breaking sector and $T$ measure
the amount of  custodial symmetry breaking. The vacuum polarizations of a gauge boson can be expanded around the zero momentum:
\beq
\Pi _{a,a'} (p^2 ) = \Pi _{a,a'} (0) + p^2 \Pi '_{a,a'} (0) +  \cdots ,
\eeq
 and the $S$ and  $T$ parameters are defined in the following way:
\begin{eqnarray}
S = 16 \pi \cdot (\Pi '_{33} (0)-\Pi '_{3Q}), \qquad \qquad T = \frac{4 \pi}{C_W^2 S_W^2 M_Z^2} (\Pi _{11} (0) - \Pi _{33} (0)).
\label{ST}\end{eqnarray}

The vacuum polarization of gauge bosons are related to propagators from the $y=L$ brane to the $y=L$ brane~\cite{RS},
\begin{eqnarray} \Pi _{11} (p) &=& g_2^2 \left( {\frac{{v^2 }}{4}} \right)^2 \left\{ {G_{p }^{( +  + )} (L,L) - G_p^{( +  + )(0)} } \right \},
\label{W11}\end{eqnarray}
\begin{eqnarray}
 \Pi _{33} (p) &=& (g_2^2  + g_{B}^2 )\left( {\frac{{v^2 }}{4}} \right)^2 \left( {G_{p}^{( +  + )} (L,L) - G_p^{( +  + )(0)} } \right) \nonumber \\
& + &  g_{Z^\prime}^2 \left( {\frac{{v^2 }}{4}} \right)^2 (\frac{1}{3} - \sin ^2 \theta )^2 G_{p}^{( -  + )} (L,L,), \label{W33}\end{eqnarray}
where $\theta$ in the above equation is the mixing angle between the $T^8$ and the $U(1)_X$ gauge bosons.  In the above calculations the
zero mode contributions have already been subtracted since we only need to integrate out the heavy KK modes. The Green's functions that we need can be
calculated from Eq.(\ref{pp}) and Eq.(\ref{mp}):
\begin{equation}
G_{p = 0}^{( +  + )} (L,L) - G_p^{( +  + )(0)}  = \frac{1}{3}\pi ^2 R^2, \qquad \qquad G_{p =0}^{( -  + )} (L,L) = \pi ^2 R^2.
\end{equation}

The gauge coupling constants after the symmetry breaking and the mixing angle $\theta$ are related to the fundamental couplings in
the following way:
\begin{equation}
g_{B}  = \frac{{\sqrt 3  g_2 g_{x} }}{{\sqrt { 3  g_2^2  + g_{x}^2 } }}, \qquad \qquad g_{Z^\prime}  = \sqrt { 3 g_2^2  + g_{x}^2 }, \qquad \qquad
\sin^2 \theta = \frac{g_{x}^2}{3 g_2^2  + g_{x}^2 }.
\end{equation}
The $U(1)_X$ coupling  $g_x$ is fixed by the measured $SU(2)_W$ coupling and the hypercharge coupling. In this way the
coupling for the heavy $Z^\prime$ is also fixed. The numerical values are $g_x=0.37$, $g_{Z^\prime}=1.18$ and $\sin^2 \theta = 0.1$. Substituting
Eq.(\ref{W11}) and Eq.(\ref{W33}) back into Eq.(\ref{ST}) we obtain
\begin{eqnarray}
T =  - \frac{1}{\alpha }\frac{{v^2 }}{4}\{ g_B^2  \cdot \frac{1}{3} + g_{Z^\prime}^2  \cdot (\frac{1}{3} - \sin ^2 \theta )^2 \} \pi ^2 R^2  \simeq
-40.0 \cdot v^2 R^2,
\end{eqnarray}
\begin{eqnarray} S =  - v^4 \{ \frac{{g_2^2  + g_B^2 }}{{45}} + \frac{{g_{Z^\prime}^2 }}{3} \cdot (\frac{1}{3} - \sin ^2 \theta )^2 \} \pi ^5 R^4
\simeq -11.5 \cdot v^4 R^4.
\end{eqnarray}
Sine the zero modes of our light fermions are flat in the fifth dimension, there are no additional contribution to the $S$ parameter as
explained in \cite{RS}. As we can see, the contribution to the $S$ parameter is almost negligible if we take $1/R$ of order a few $\mbox{TeV}$.
The constraint from the $T$ parameter is more stringent. The current PDG fit requires $T= - 0.17 \pm 0.12$ and with $S=0$, it gives a lower
bound $T
> -0.15$. For $v=246$ GeV, this condition requires the inverse radius to satisfy $1/R > 4 \mbox{ TeV}$.

Similarly, we can calculate the extra contribution to $Z \rightarrow b_L b_L$ by integrating out the heavy gauge bosons. Since our fermions live
in the bulk and the Higgs lives on the $y=L$ boundary, we need Green's functions that propagate from an arbitrary position in the extra
dimension to the $y=L$ boundary. When the Higgs field gets a VEV, it mixes the zero mode of the $Z$ gauge field with the KK modes of $W^3$, $B$
and $Z^\prime$.  Here $Z^\prime$ corresponds to the massive combination of the $T^8$ generator of $SU(3)_W$ and the $U(1)_X$ that is broken by
the $y=0$ boundary. The net contributions from integrating out heavy $SU(2)$ gauge bosons $W^3_{\mu}$ and heavy hypercharge gauge bosons
$B_{\mu}$ are zero. There is only a contribution from the heavy $Z^\prime$ gauge bosons:
\begin{eqnarray}
 \frac{\delta g_{bL}}{g_{bL}}  &=& \frac{{g_{Z^\prime}^2 v^2 }}{2} \cdot \frac{Q_{Z^\prime}^{bL}Q_{Z^\prime}^H}{Q_Z^{bL}}
 \cdot \int_0^L {dyf_0^2 (y)} G_{p = 0}^{( -  + )} (y,L) \nonumber \\
 &=& \frac{{g_{Z^\prime}^2 v^2 }}{2} \cdot \frac{Q_{Z^\prime}^{bL}Q_{Z^\prime}^H}{Q_Z^{bL}}
 \cdot \int_0^{\pi R} {y dy}. \label{zbb}
 \end{eqnarray}
 In the above equation, $f_0(y)$ is the profile of the left handed bottom quark $b_L$ in the extra dimension, and  $Q_{Z^\prime}^H$,  $Q_{Z^\prime}^{bL}$, and $Q_{Z}^{bL}$ are the
$Z^\prime$ and $Z$ charges of the Higgs and the left-handed bottom quark:
\beq
Q_{Z^\prime}^H  = \frac{1}{6} - \frac{1}{2}\sin ^2 \theta , \qquad Q_{Z^\prime}^{bL}  = \frac{1}{6} + \frac{1}{6}\sin ^2 \theta , \qquad  Q_Z^{bL}  =  -
\frac{1}{2} + \frac{1}{3}\sin ^2 \theta _W .
\eeq
Evaluating Eq.(\ref{zbb}), we get $\delta g_{bL} /g_{bL}  =  - 0.17 \cdot v^2 R^2$. LEP data requires that $ \delta g_{Lb} / g_{Lb}
< 0.1 \%$~\cite{LEP}, which gives a constraint of  $1/R > 3.2 \mbox{ TeV}$.

The 4-fermion interactions mediated by KK modes of $Z^\prime$ also give some constraints. The most stringent constraint is from the composite
scale $\Lambda^+_{RL}(eeuu) > 23.1 \mbox{ TeV}$~\cite{Cheung:2001wx}. It is required that:
\begin{eqnarray}
 \left( g_{Z'}^2 Q_{Z'}^{eR}Q_{Z'}^{uL} {\int_0^{\pi R} {dy_1 \int_{y_1 }^{\pi R} {dy_2 } f_0^2 } (y_1 )f_0^2 (y_2 )G_{p = 0}^{( - + )} (y_1 ,y_2 )}
\right)^{ - 1}  > \frac{\Lambda^{+ 2}_{RL}}{4 \pi} ,
\end{eqnarray}
where
\beq Q_{Z^\prime}^{eR}  = \frac{1}{3} - \sin ^2 \theta , \qquad Q_{Z^\prime}^{uL}  = \frac{1}{6} + \frac{1}{6}\sin ^2 \theta \eeq
are the $Z^\prime$ charges for the right-handed electron and the right-handed up quark. The
contributions from all of the KK modes of the $Z^{\prime}$ gauge bosons are included. This puts a constraint of $1/R
> 2.0 \mbox{ TeV}$.

\section{Quirk Phenomenology}

 In order to conduct an analysis of the phenomenology, we need to have some information about the various scales in the model.  We will assume the top quirks $X_T$ and $T_R$ to be the lightest fermion in the
infracolor gauge sector. The masses of the other quirks are controlled by the size of the extra dimension and the brane mass parameters and for
simplicity we can take them to be around $10$ TeV. The scale, $\Lambda_C$, where the $SU(3)_C$ gauge coupling blows up is around $100$ MeV while
the quirk phenomenology depends sensitively on the scale $\Lambda_{I}$ where the $SU(3)_{I}$  infracolor gauge coupling blows up. The bulk
$SU(6)$ gauge dynamics  would tend to set the couplings $\alpha_{I}(1/R)$ and $\alpha_{I}(1/R)$ equal, but they could  be different due to the
boundary gauge kinetic terms at $y=0$.

To get an upper bound for the  $\Lambda_{I}$ scale, we assume $\alpha_{I}(1/R)\leq 2$ so that a perturbative expansion still works. Running from
the scale $1/R$ to the top quirk mass $M_{X}$, only the infracolor gluons and the top quirks contribute.
\begin{eqnarray} \Lambda  &=& \frac{1}{R} \exp \left[ - \frac{{2\pi }}{{\alpha _I (1/R)}}\frac{3}{{31}} \right]
\end{eqnarray}
To run below  $M_{X}$, we can decouple the top quirk so the scale $\Lambda_{I}$ is related to the scale $\Lambda$ in the following way:
\begin{eqnarray} \left( \frac{\Lambda _{I} }{M_{X} } \right)^{11}  = \left( {\frac{\Lambda }{M_{X}}} \right)^{31/3}
\end{eqnarray}
In Table \ref{scale} we show some different values for $\alpha_I(1/R)$ and the corresponding value of $\Lambda_I$, assuming $1/R= 10$ TeV and
$M_{X}= 1$ TeV.
\begin{table}[!htb]
\begin{center}\begin{tabular}{|c|c|c|c|c|} \hline  $\alpha_I(1/R)$ & 0.24 & 0.036 & 0.028
 & 0.023  \\\hline $\Lambda_{I}$ & $800$ GeV &
MeV & $10$ keV & $100$ eV
\\\hline L & $0.246 \mbox{\AA}$ & $0.1 \mu\mbox{ m }$ & $ 1.0 \mbox{mm}$ & $10 \mbox{ m }$\\\hline
\end{tabular}\end{center} \caption{Different scenarios for the infracolor gauge coupling constant $\alpha_I(1/R)$
and infracolor scale $\Lambda_{I}$ for $1/R= 10$ TeV and $M_{X}= 1$ TeV.}\label{scale}\end{table}

Quirk phenomenology in the detector depends sensitively on the infracolor scale $\Lambda_{I}$, there are several very different possible cases
\cite{quirk,FP,Cheung,Harnik}.

1) MeV $ <\Lambda_{I} < M_{X}/ \mbox{few}$: Microscopic strings. The quirk---anti-quirk pair annihilates mainly  into hidden sector glueballs,
but also into two photons as well as two light SM quarks or two leptons.  The displaced leptons may provide the easiest search strategy.

\vspace{10pt}

 2) $10$ keV $ <\Lambda_{I} < $ MeV: Mesoscopic strings.
In this range, quirk---anti-quirk  pairs can form mesoscopic strings (flux tubes) which appear as single particle tracks
in the detector. Their ionization is different from SM particles and  the mass of the bound state differs event-by-event. Since the bound state is neutral it will not bend in the magnetic field of the detector.

\vspace{10pt}

 3) $100$ eV $ <\Lambda_{I} < 10$ keV: Macroscopic strings.
The two quirks form macroscopic strings. The string interaction attracts the quirk and the anti-quirk towards  each other and they leave two
separate anomalously curved tracks in the detector.

 Since our quirks are uncolored. Their main production channel for $q\bar q \to X\bar X$ in LHC is through s-channel exchange of photon and $Z$ gauge bosons:

\begin{eqnarray}\frac{{d\sigma }}{{dt}}&=& \frac{{e^4 Q_X^2 }}{{64\pi s^4 }}
 \left((M_X^2  + \hat s)^2  + 2t^2  - 2(M_X^2  - \hat s)t \right )
\nonumber \\ &&  \{ 8Q_q^2 + \frac{{s^2 (8Q_q^2 SW^4  - 4Q_q SW^2  + 1)}}{{CW^4 (M_Z^2  - \hat s)^2 }} + \frac{{\hat s(M_Z^2 - \hat s)4Q_q (1 -
4Q_q SW^2 )}}{{CW^2 (M_Z^2  - \hat s)^2 }} \}
\end{eqnarray}

In the above equations, $\hat s$ and $\hat t$ are the Lorentz-invariant Mandelstam variables at the parton level. At the LHC, the two quarks come from two protons, so we need to integrate the differential cross section over parton distribution
functions, $P_{a/P} (x_a,Q^2 )$, in each proton.  The total cross section is
\begin{eqnarray}
\sigma = \int_{x_a\min }^1 {dx_a} \int_{x_b\min }^1 {dx_b} \int_{\hat{t} \min}^{\hat{t} \max} {d\hat{t}} P_{a/P} (x_a,Q^2 ) P_{b/P} (x_b,Q^2
)\frac{{d\sigma }}{{d \hat t}}
\end{eqnarray}
There are several schemes for the definition of $Q^2$ and we take $Q^2=-\hat t$. With the threshold constraint $\hat{s}=x_a x_b s < 4M_{X}^2$,
we can fix the lower bound of the two fraction parameters $x_a$ and $x_b$: $x_{a \min} = \frac{4M_X^2}{s}$ and $x_{b \min}=
\frac{4M_X^2}{s}\frac{1}{x_a}$. The upper and lower bounds of ${\hat t}$ can be determined from:
\begin{eqnarray}
\hat t_{\min }  &=&  - \frac{1}{2}\left( {(\hat s - 2M_{X}^2 ) + \sqrt
{(\hat s - 2M_{X}^2 )^2  - 4M_{X}^4 } } \right) \\
\hat t_{\max }  &=&  - \frac{1}{2}\left( {(\hat s - 2M_{X}^2 ) - \sqrt {(\hat s - 2M_{X}^2 )^2  - 4M_{X}^4 } } \right)
\end{eqnarray}

\begin{figure}[htb]
\begin{center}
\includegraphics[width=0.8\hsize]{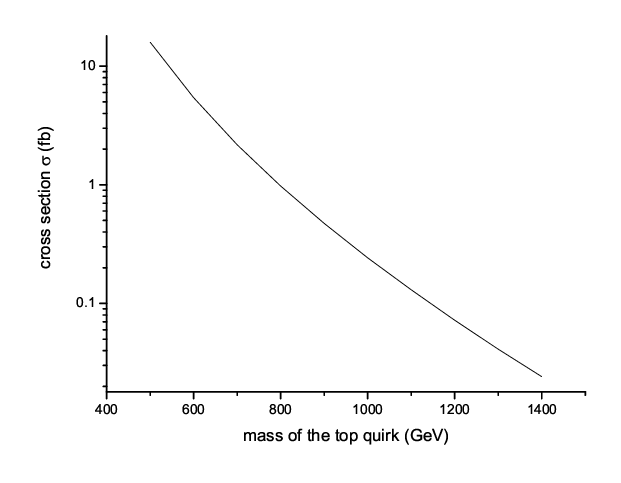}
\end{center}
\caption{Total cross-sections vs. mass of top quirk \label{fig:quirk}}
\end{figure}

The cross section for pair production of quirks at the LHC is shown in Fig.~\ref{fig:quirk}. The LHC will run at  $s=(14 \mbox{ TeV})^2$ and we
will take a luminosity of $3 \cdot 10^{2} \mbox{ fb}^{-1}$. For a top quirk mass $M_{X}=800$ GeV, about one hundred events with quirk pairs can
be produced.

\section{Conclusion}

 In this paper, we have displayed a quirky little Higgs model and used a color-neutral top quirk to cancel the quadratic divergence from
the top quark loop. The top quirk and top quark are related by an $SU(6)$ bulk gauge symmetry in which their respective confining gauge groups are embedded.  The Higgs in this model is a pseudo-Nambu-Goldstone boson and its
mass parameter is protected by an $SU(3)_W$ symmetry. The collective breaking of the little Higgs mechanism occurs on two separate branes, which
leads to finite results for the Higgs mass. Since the mass spectrum is mainly determined by the radius of extra dimension, precision electroweak
tests only put stringent constraints on $1/R$. This is quite different from  the original little Higgs model, there the mass of the heavy $Z^\prime$ gauge boson is determined by the
scalar VEV $f$, which puts tight  constraints on the parameter $f$. However, in our model, there is no sensitivity to the parameter $f$.
Here we required $f$ to be around $800$ GeV so that the magnitude of the negative radiative correction from the top quark and top quirk loops could be
larger than the gauge contribution. This allows for radiative electroweak symmetry breaking in our model.

In our model, the top quirk is color-neutral and its
main production mechanism is through quark annihilation. For quirks with a mass less than $1 \mbox{TeV}$, there are large numbers of events with quirk
pair production at the LHC. The signature of quirk pairs in the detector depends strongly on the infracolor gauge coupling. For stronger
values of the infracolor coupling, there will be large SM backgrounds and the signal may be very hard to differentiate, but for weaker values
there are long strings between the quirks which makes them much easier to find.

\section*{Acknowledgments}
We thank Z. Chacko, R. Harnik, M. Luty, and J. Thaler for helpful discussions.
This work was supported in part by U.S. DOE grant No. DE-FG02-91ER40674.

\end{document}